# Detecting and Correcting Gain Jumps in TES Microcalorimeters

Thomas A. Baker, Daniel T. Becker, Joseph W. Fowler, Mark W. Keller, Daniel S. Swetz, and Joel N. Ullom

*Abstract*—**Arrays of microcalorimeters based on transition-edge sensors (TESs) are being actively deployed to laboratories all over the world. A TES microcalorimeter array produces very large quantities of data and users of these devices have varying levels of experience, so it is important to provide robust software for data acquisition and analysis that can function with minimal user supervision. This software should be capable of addressing common phenomena that can adversely affect spectrum quality. Gain jumping is one such phenomenon that is characterized by abrupt changes in the gain of a device. Left unaddressed, gain jumps can degrade spectra by introducing false peaks. We are not aware of any previously published methods for resetting gain jumps during data acquisition or existing algorithms for correcting data that is degraded by gain jumps. We have developed automated methods for detecting and correcting gain jumps in gamma-ray TES microcalorimeters. We present a procedure for resetting gain jumps during a live data acquisition that involves briefly driving the TES into its normal state using the bias current. We also describe an algorithm for locating gain jumps and identifying unique gain states within existing microcalorimeter data. Finally, we provide a possible approach for correcting gain jumps after they have been identified.**

*Index Terms*—**Data analysis, gamma rays, detectors, spectroscopy.**

## I. INTRODUCTION

**M**ICROCALORIMETERS based on transition-edge sensors (TESs) have had great success as spectrometers thanks to their high energy resolution compared to commercial alternatives [1]. These detectors have a wide range of applications in areas such as nuclear materials analysis [2], [3], [4], exotic atom physics [5], [6], [7], particle physics [8], and astronomical measurement [9], [10], [11]. Microcalorimeter arrays containing hundreds of individual detectors fabricated by our team are being actively deployed to increasing numbers

Manuscript received 25 September 2024; revised (date). Date of publication (date); date of current version (date). The information, data, or work presented herein was funded in part by the Advanced Research Projects Agency-Energy (ARPA-E), U.S. Department of Energy, under Award Number DE-AR0001693, and also by the Office of Nuclear Energy, U.S. Department of Energy, under the MPACT Program. The views and opinions of authors expressed herein do not necessarily state or reflect those of the United States Government or any agency thereof. (*Corresponding author: Thomas A. Baker*).

Thomas A. Baker, Joseph W. Fowler, and Daniel T. Becker are with the Department of Physics, University of Colorado Boulder, Boulder, CO 80309 USA (e-mail: thomas.baker-2@colorado.edu).

Mark W. Keller and Daniel S. Swetz are with the National Institute of Standards and Technology, Boulder, CO 80305 USA.

Joel N. Ullom is with the Department of Physics, University of Colorado Boulder, Boulder, CO 80309 USA, and also with the National Institute of Standards and Technology, Boulder, CO 80305 USA.

of users with limited experience in their operation, and as such, it is important to automate not only the acquisition but also the analysis of data from these devices. Whenever possible, analysis of microcalorimeter pulse data should include corrections for common errors that can degrade the resolution of the devices or the quality of the spectra they produce.

The raw data produced by a TES microcalorimeter consists of a set of pulse records and timestamps. Pulses are produced when a photon or energetic particle deposits energy within the detector, resulting in a brief increase in temperature and a change in the current through the detector. The height of the pulse corresponds to the energy deposited in the detector [12], although the exact relationship between height and energy can drift in time due to slight fluctuations in detector properties or environmental conditions [13]. Often, these fluctuations also produce a change in the baseline current through the detector, and this baseline level can be used to remove drift from the data. Existing software for microcalorimeter data analysis performs this correction [14].

An error that is not reliably detected or corrected by existing data analysis techniques is a phenomenon known as "gain jumping." This occurs when the device gain abruptly shifts between discrete levels in a manner that cannot be corrected using a correlation to the device baseline. Previous observations have suggested that gain jumps are triggered by detector absorption events, but the extent to which this holds true is uncertain. The different gain levels are thought to correspond to distinct operating states of the TES, with each state corresponding to a different number of kinematic vortex streets [15] and/or phase slip lines [16] within the TES. We are not aware of any prior research published on this hypothesis, however. When ignored, gain jumps can cause serious degradation of spectra, so it is important to either prevent them during data acquisition or correct them when processing the data.

Gain jumps have previously been observed in data from both X-ray and gamma-ray microcalorimeters. Techniques exist for resetting the gain state of X-ray devices to a ground level during data acquisition. These techniques were not published, nor were they extended to gamma-ray devices. Furthermore, no reliable techniques were developed to detect and correct gain jumps in existing data for either X-ray or gamma-ray devices. In this paper, we summarize our work in reversing gain jumps in gamma-ray microcalorimeters during live data acquisition. We also describe an offline algorithm for identifying gain states in microcalorimeter data and suggest an automated data-processing approach that maintains spectral quality while minimizing discarded data.

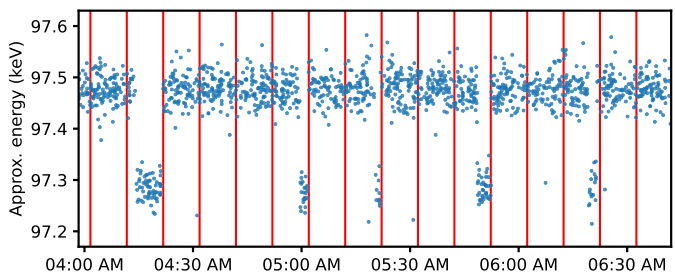

Fig. 1. Effect of spiking the bias at 10-minute intervals on a single device that was susceptible to gain jumping. The line for $^{153}$Gd at 97.43 keV is shown. Points represent approximate pulse energies and vertical red lines indicate bias spikes.

## II. EXPERIMENTAL SETUP

We used a $^{153}$Gd calibration source and a 96-detector microsnout array (described in [17]) to collect data for developing and testing our gain jump detection and correction techniques. Each detector consists of a Sn absorber connected thermally to a Mo/Cu bilayer TES on a silicon nitride membrane. Each absorber has dimensions $1.25 \times 1.25 \times 0.5$ mm. The devices were voltage-biased and were read out using microwave SQUID multiplexing [18]. The $T_c$ for these detectors was $118 \pm 5$ mK. Faulty wiring for our test cryostat prevented the use of a quarter of the detectors, and some detectors were unusable for other reasons, so we were limited to 68 usable detectors. We operated the array at a bath temperature of 85 mK and selected a bias current that placed the detectors in the transition region at about 10% of their normal resistance. We selected this bias point in order to maximize the rate of gain jumping while otherwise preserving detector stability. Bias points lower than 10% resulted in a higher rate of gain jumping, but also led to worse detector stability. About 38 of the detectors were subject to gain jumping under these conditions. We performed baseline-correlated drift correction [14] on the data from this array before performing further analysis.

We also generated synthetic data to develop and test the offline gain jump detection algorithm. The data consists of a list of pulse heights for a single spectral line containing gain jumps. The pulse heights were centered on various gain state levels, with abrupt shifts from one level to another at random time intervals. We used a Gaussian line shape, a uniform background, and a range of different values for gain state separation, probability of jumping to each state, Gaussian line width, background level, number of unique gain states, and total event count. This allowed us to test our algorithm on data with a wide range of characteristics.

## III. RESET METHOD

Our method for resetting the gain state of the TES is to briefly drive the device normal using the bias current. We used a bias current that placed the detector well within the normal branch, and gradually returned to the operating point using small steps over a period of about 3.5 seconds. This slow return was chosen because we found that rapid changes in the bias produced less consistent device states. We refer to the reset operation as a "bias spike." Rather than detect gain jumps in a live data stream, we used bias spikes simultaneously on all channels at 10-minute intervals, with the hope that some of these bias spikes would occur by chance at times when a device was in a secondary gain state. Because each bias spike lasted 3.5 seconds, the dead time was only about 0.6% of the total acquisition time.

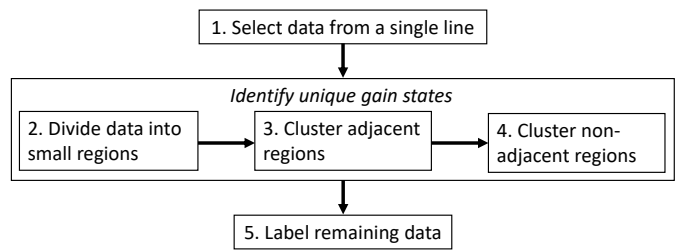

Fig. 2. Schematic showing the main steps within our offline algorithm for identifying gain states in microcalorimeter data.

We collected data for 48 hours with our bias spike running for the first 20 hours. We found that our method caused the detectors to return to a consistent gain state after each spike (Fig. 1). Visual inspection of the plotted data suggested that when a detector was already in the desired gain state, or when a detector was not susceptible to gain jumping, the bias spike did not affect the quality of the data. We also observed that the gain state produced by the bias spikes is not necessarily the state in which the device would spend the most time in the absence of bias spikes. More investigation would be necessary to confirm this.

## IV. OFFLINE DETECTION AND CORRECTION

### A. Detection Algorithm

The algorithm for detecting unique gain states in microcalorimeter data requires a dataset consisting of an array of individual pulse height values in chronological order and a pulse height range that describes the location of a reference spectral line. Note that, in the following discussion, all mention of statistical properties refers to data from this array of pulse height values; for example, "noise" refers to variation in pulse height. The algorithm requires no prior knowledge about the energy resolution of the detector nor the locations, separations, and/or number of gain states. We make a few assumptions about the data. First, we assume that the specified height range contains only a single spectral line. We also assume that all continuous drift has been removed, so that the line is piecewise-constant with discontinuities at the locations of the gain jumps. We treat the line shape as Gaussian for mathematical convenience, and we do not expect small deviations from that shape to significantly impact the performance of the algorithm.

The algorithm consists of five main steps, depicted schematically in Fig. 2. The first step is to select the data within the spectral line. To do this, we first cut all values that fall outside of the specified pulse height range. Then we calculate a running median and a running median absolute deviation for the remaining values. We use the median of this running median absolute deviation to estimate the standard deviation of the Gaussian line shape, and then we cut all values that fall more than four standard deviations away from the running median.

The second step is to divide the data into regions. We first estimate the standard deviation of the noise in the selected data. Then we calculate a smoothed version of the data by applying two passes of a bilateral filter [19] with Gaussian kernels to a copy of the data. The bilateral filter is an edge-preserving filter, and this slightly improves the accuracy in this step. In the first pass, the spatial kernel has a standard deviation of four samples, and the range kernel has a standard deviation equal to four times the estimated standard deviation of the noise. In the second pass, the spatial kernel has a standard deviation of two samples, and the range kernel has a standard deviation equal to the estimated standard deviation of the noise. After this, we identify all inflection points in the resulting line. These points are candidates for gain jumps and divide the data into a set of small regions.

The third step is to cluster the regions. We use a variant of agglomerative clustering [20] to accomplish this. The algorithm repeatedly cycles through the regions and compares each region to its neighbors to determine whether they are part of the same gain state, and if so, it merges them. The main steps in the comparing/merging process are as follows:

a) For a given region $R_i$, calculate the standard deviations $\sigma_{i\pm 1}$ of the neighboring regions $R_{i-1}$ and $R_{i+1}$. Then calculate two possible values for the standard error of the mean for region $R_i$ under the assumption that it has the same standard deviation as one of its neighbors, according to the formula $s_{i\pm} = \sigma_{i\pm 1}/\sqrt{N_i}$, where $N_i$ is the size of region $R_i$.

b) Calculate a threshold separation for each of the two regions by taking a user-defined multiple of $s_{i\pm}$ associated with that region.

c) Calculate the distance from region $R_i$ to each of the neighboring regions, where the distance is defined as the difference between the means of the regions minus the threshold separation.

d) If the number of samples in region $R_i$ is less than a user-specified threshold value, or if the distance between region $R_i$ and either $R_{i-1}$ or $R_{i+1}$ is negative (i.e. region $R_i$ falls within the threshold separation), then merge region $R_i$ with the closer of regions $R_{i-1}$ and $R_{i+1}$.

This process is repeated many times for all regions, but with reduced threshold values for the first few iterations. It finishes when the remaining regions do not satisfy the threshold requirements for merging. The region boundaries are the estimated locations of gain jumps.

In the fourth step, we identify which regions correspond to the same gain state. We cluster the regions based on distance between their means, where the distance between two regions $R_i$ and $R_j$ is specified as a multiple of the greater of $\sigma_i$ and $\sigma_j$ plus the greater of $s_i$ and $s_j$, where $s_i = \sigma_i/\sqrt{N_i}$. Clustering continues until all distances are greater than a user-specified threshold value.

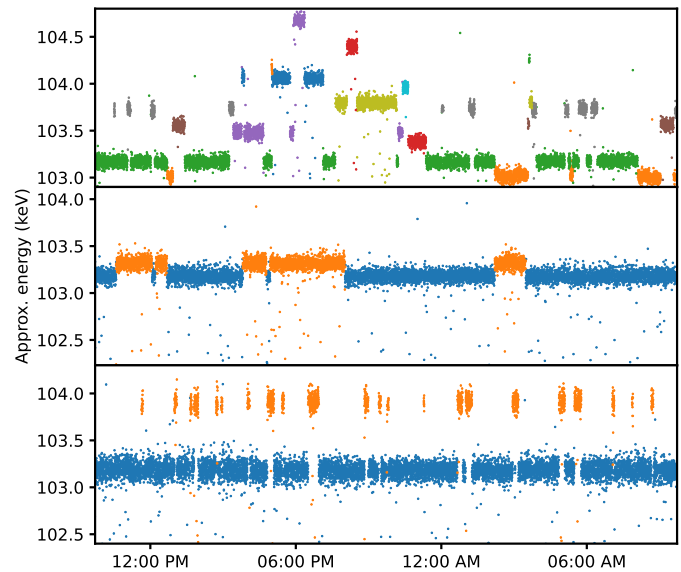

Fig. 3. Approximate pulse energies from three different detectors, color-coded based on the gain states identified by our offline detection algorithm. Vertical axis scale is the same for all three plots, making apparent the differences in linewidth and state spacing. The first plot contains 13 gain states, so some colors were reused.

In the fifth and final step, we assign gain states to the values that were removed in step 1, by using the nearest known neighbor according to sample number.

After performing these steps, the algorithm returns an array of labels, one for each of the original data values, that identify the gain state corresponding to each value.

### B. Testing the Algorithm

We tested our offline detection algorithm on synthetic data with known gain jump amplitude and timing. The algorithm reproduced the expected solution to within acceptable limits, and the overall accuracy was comparable to what a human researcher might obtain through manual analysis.

After this, we applied our algorithm to real data containing gain jumps with unknown amplitude and timing. We took the dataset from our bias spike test and selected the data from the final 28 hours when the bias spike was disabled. This time period contained about 50,000 events per detector. We applied our detection algorithm to each of the 68 detectors, and provided a pulse height range that corresponded to approximately 101 keV to 105 keV and contained a single spectral line at 103.18 keV. This line is slightly weaker than the line at 97.43 keV, but it is the highest line in the spectrum for $^{153}$Gd (not counting the very weak line at 172.85 keV) and can be reliably located in the presence of gain jumps. For parameters, we used a minimum region size of 5 samples, a threshold separation of 5.0 standard errors of the mean, and a state-clustering threshold of 1.0. The parameters were the same for all detectors. Some examples of data that was labeled by the algorithm are provided in Fig. 3.

After identifying the unique gain states and the locations of the jumps between states, there are multiple options for correcting the data. One solution is to select data from the most populated gain state and discard the remaining data. This option may be beneficial when the other gain states are brief and contain very few samples. Another option is to keep all of the data, but to divide it into separate groups corresponding to unique gain states. These groups may then be processed separately as if they were different detectors, or discarded when the number of samples within the group is too small. We chose this approach for our analysis. Fig. 4 provides a spectrum before and after correcting gain jumps within the data. The result contains no sign of duplicate peaks and is a high-quality spectrum, showing that our approach was successful.

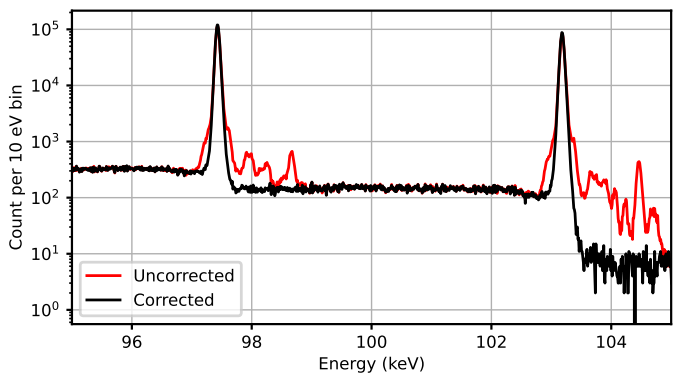

Fig. 4. Spectral lines for $^{153}$Gd at 97.43 and 103.18 keV before and after applying our correction method using the 103.18 keV line. The uncorrected spectrum was produced by separately calibrating and adding the results from 68 separate detectors. To produce the corrected spectrum, we divided the data from each detector into separate gain states, performed energy calibration on each state separately, and summed the results. A total of 113 individual gain states were calibrated.

A tempting alternative approach for correcting gain jumps is to linearly rescale the data in each gain state region using a piecewise-constant fit to the selected spectral line. This would avoid the need to repeat the energy calibration on each region separately. However, this solution fails because gain jump spacing has a nonlinear dependence on energy, so even if the selected spectral line has been corrected, other lines may not be corrected. The approach we took is robust to this issue because ordinary energy calibration does not assume linear dependence and can adjust different regions of a spectrum independently from one another.

## V. Discussion

The effectiveness of our bias spike method for resetting gain jumps supports the hypothesis that gain jumps occur within the TES itself. It also shows potential for future applications, but without further development its areas of application are limited. In its current implementation, bias spikes occur at fixed intervals, and a detector could remain in an undesirable gain state for the duration of that interval before a bias spike resets it. One solution would be to apply bias spikes immediately after every gain jump, but this would require developing an online jump detection algorithm. Another tempting solution is to keep the bias spikes at regular intervals, and simply to correct the remaining gain jumps using our offline correction method, but we think this approach could give worse results than simply using the offline correction by itself. This is because the bias spike correction results in many brief excursions to other gain states, and such excursions are more difficult to identify accurately. Ultimately, we believe that the best approach is to only use our offline correction for gain jumps and not to use bias spikes at all, except in scenarios where offline correction is impossible (for example, when no well-isolated spectral line exists).

Our offline algorithm for detecting gain jumps has worked well on our test data, and we expect it to be a valuable tool for other datasets. Nevertheless, our algorithm has certain limitations that could point to areas for future improvements. One limitation is that the algorithm requires an isolated spectral line, and in some scenarios such a line may not exist. Furthermore, very large gain jumps may cause even a well-isolated line to interfere with other lines in the spectrum, but our current algorithm cannot reliably identify and correct these types of gain jumps. Our algorithm also requires that the reference spectral line will reveal all gain jumps that occur, but it is possible that the nonlinear energy dependence of gain state separation could cause gain jumps to vanish in one spectral line but not others, in which case the algorithm would not detect them. Finally, due to the fact that a single spectral line is used for reference, the algorithm cannot locate a gain jump with greater precision than the separation between consecutive data points within that line. Consequently, data points from other spectral lines that occur in the close vicinity of a gain jump may be assigned to an incorrect gain state, so it may be necessary to discard data near each gain jump. These limitations are likely not an issue for most applications, but we could investigate some strategies to mitigate them if they become a concern.

Our analysis of the impact of our algorithm is primarily qualitative. A quantitative analysis is difficult because the degradation caused by gain jumps is not always quantifiable. A spectrum produced from data containing gain jumps will often show the same resolution as a clean spectrum, with the only noticeable difference being the appearance of additional peaks in the spectrum. The negative effect of these additional peaks could be a false positive for the presence of an element, an inaccurate estimate of concentration, or some other negative outcome depending on the particular details of the experiment. Thus, the level of impact of a gain jump may be completely unrelated to its height or amplitude.

One of the advantages of our offline gain jump detection algorithm is that it adapts to differences in linewidth, separation, and number of gain states without any need for tuning the algorithm parameters. The parameters have no reference to the absolute scale of the data, so the same parameters can be used for very different datasets. Nevertheless, some tuning may still be necessary based on the goal of the analysis. For example, if the goal is to generate a high-resolution spectrum, it may be beneficial to detect and correct slight continuous drift as if it were composed of gain jumps, in which case less tolerant parameters would be appropriate. But if the user wishes to identify the gain states themselves without any false positives, a more tolerant set of parameters could be used to exclude continuous drift.

Overall, our algorithm shows great promise for use in both current and future systems. Although some areas could be developed further, the current version is already sufficient for useful application to experimental data. Our tests so far have been limited to gamma-ray microcalorimeters but we expect the algorithm to work well on data from detectors for other energy ranges.